\documentclass{article}[12pt]
\usepackage{natbib}
\usepackage{amsmath}
\usepackage{graphicx}

\begin{document}
\title{The influence of the radiative non-symmetric ion-atom collisions on the
stellar atmospheres in VUV region}
\author{V.A.Sre{\' c}kovi{\' c}$^{1}$, A.A.Mihajlov$^{1}$, Lj.M.Ignjatovi{\' c}$^{1}$ and M.S.Dimitrijevi{\' c}$^{2}$\\\\
$^{1}$University of Belgrade, Institute of physics, P.O. Box 57, 11001,\\ Belgrade, Serbia\\
 $^{2}$Astronomical Observatory, Volgina 7, 11060 Belgrade
Serbia\\ and \\Observatoire de Paris, 92195 Meudon Cedex, France}
\maketitle

\begin{abstract}
The aim of this work is to draw attention to the processes of
radiative charge exchange in non-symmetric ion-atom
collisions as a factor of influence on the opacity of stellar
atmospheres in VUV region. For that purpose calculations of the
spectral absorption coefficients for several ion-atom systems, namely:
He + H$^{+}$ and H + X$^{+}$, where X = Na and Li have been performed.
On chosen examples it has been established that the examined processes
generate rather wide molecular absorption bands in the VUV region, which should be
taken into account for the interpretation of data obtained from laboratory measurements
or astrophysical observations. In this paper the potential significance is discussed
of the considered radiative processes for DB white dwarfs and solar atmospheres,
as well as for the atmospheres of the so-called lithium stars.
\end{abstract}

%\keywords{atomic processes --- stellar atmospheres}

%%%%%%%%%%%%%%%%%%%%%%%%%%%%%%%%%%%%%%%%%%%%%%%%%%%%%%%%%%%%%%

\section{Introduction}

A series of previous papers studied the influence of the processes of radiative charge
exchange in symmetric
$H(1s)+H^{+}$ and $He(1s^{2}) + He^{+}(1s)$ collisions and corresponding
photo-association/dissociation processes
on the opacity of stelar atmospheres. In the hydrogen case, these processes were shown to be
important for the atmospheres of the Sun and some DA white dwarfs \citep{sta94}; in the
helium case, for the atmospheres of some DB and DA white dwarfs \citep{mih92b, sta94}. Thus,
the mentioned papers made it clear that at least symmetric ion-atom radiative collisions
play a significant role in stellar atmospheres. But the question whether some non-symmetric
ion-atom radiative processes can also influence the optical characteristics of the considered
stellar atmospheres is still open. A detailed study of such non-symmetric processes in
connection with the stellar atmospheres would require very extensive research, and remains a
task for the future. The aim of this short article is to point out at least some objects where
such processes could be of interest, and to show the possible ways of describing their
influence. For this purpose it was natural to start from the same DB white dwarfs'
and the solar atmosphere (which were considered in previous papers, since
adequate models exist for them). In the case of DB white dwarfs we mean models presented
in \citet{koe80}, as well as in \cite{ber95}. The necessary models of the solar atmosphere were
described in \citet{ver81} and \citet{mal86}, and in \citet{fon06, fon09}.

The composition of the mentioned models and the previous results for the ion-atom symmetric
radiative process suggest that in the considered atmospheres the following non-symmetric
absorption processes have to be taken into account
%%%%%%%%%%%%%%%%%%%%%%%%%%%%%%%%%%%%%%%%%%%%%%%%%%%%%%%%%%%%%%%%%%%%
\begin{equation}
\label{eq:HeH1} \varepsilon_{\lambda} + HeH^{+} \longrightarrow He^{+}
+ H,
\end{equation}
%%%%%%%%%%%%%%%%%%%%%%%%%%%%%%%%%%%%%%%%%%%%%%%%%%%%%%%%%%%%%%%%%%%%
\begin{equation}
\label{eq:HeH2} \varepsilon_{\lambda} + He + H^{+} \longrightarrow
He^{+} + H,
\end{equation}
%%%%%%%%%%%%%%%%%%%%%%%%%%%%%%%%%%%%%%%%%%%%%%%%%%%%%%%%%%%%%%%%%%%%
%%%%%%%%%%%%%%%%%%%%%%%%%%%%%%%%%%%%%%%%%%%%%%%%%%%%%%%%%%%%%%%%%%%%
\begin{equation}
\label{eq:HX1} \varepsilon_{\lambda} + HX^{+} \longrightarrow H^{+}
+ X,
\end{equation}
%%%%%%%%%%%%%%%%%%%%%%%%%%%%%%%%%%%%%%%%%%%%%%%%%%%%%%%%%%%%%%%%%%%%
\begin{equation}
\label{eq:HX2} \varepsilon_{\lambda} + H + X^{+} \longrightarrow
H^{+} + X,
\end{equation}
%%%%%%%%%%%%%%%%%%%%%%%%%%%%%%%%%%%%%%%%%%%%%%%%%%%%%%%%%%%%%%%%%%%%
where $He,He^{+},H \equiv He(1s^{2})$, $He^{+}(1s)$ and $H(1s)$ respectively, $HeH^{+}$ and $HX^{+}$ are
the molecular ions in the corresponding electronic ground states, and $X$ and $X^{+}$ -
a metal atom and its ground-state ion. In the general case, apart from these absorption
processes, the corresponding inverse emission processes should also be considered. However, it
can be shown that, under the conditions of plasma taken from the models mentioned above, the
significance of such emission processes can be neglected in comparison with other relevant
emission processes.
In this work we will consider the
absorption processes (\ref{eq:HeH1}) and (\ref{eq:HeH2}) in connection with DB white dwarfs
and solar atmospheres, and the processes (\ref{eq:HX1}) and (\ref{eq:HX2}) with $X=Na$ and
$X=Li$ - in connection with the solar and so-called lithium stars' atmospheres.

\section{Theoretical remark}

The ground and
 first excited electronic state of the considered molecular ion
($HeH^{+}$ or $HX^{+}$) will be denoted here by $|1>$ and $|2>$, and the
corresponding adiabatic potential curves - by $U_{1}(R)$ and
$U_{2}(R)$, where $R$ is the internuclear distance. By $D_{12}(R)$
will be denoted the modulus of the dipole matrix element for the
transition $|1> \rightarrow |2>$, i.e. $D_{12}(R)=|<1|{\bf D}(R)|2>|$,
where ${\bf D}$ is the operator of dipole moment of $AX^{+}$.
Keeping in mind the results of the previous papers, we will treat the
non-symmetric radiative processes (\ref{eq:HeH1}) - (\ref{eq:HX2}) within the
dipole approximation where the data about $U_{1}(R)$, $U_{2}(R)$ and $D_{12}(R)$
are sufficient for all the necessary calculations.
In accordance with \citet{mih07a} and \citet{ign09} the quantum-mechanical treatment is applied
here to the
photo-dissociative processes (\ref{eq:HeH1}) and (\ref{eq:HX1}), while the absorption charge
exchange processes (\ref{eq:HeH2}) and (\ref{eq:HX2}) are described within the semi-classical
approximation.

The efficiencies of the photo-dissociative process (\ref{eq:HeH1}) and
(\ref{eq:HX1}) are characterized here by the partial spectral absorbtion
coefficients $\kappa_{1,3}^{(a)}(\lambda)=
\sigma_{1,3}^{phd}(\lambda,T)\cdot N_{mi}$, where $\sigma_{1,3}^{phd}(\lambda,T)$
is the the photo-dissociation cross-section, and the absorption charge exchange
processes (\ref{eq:HeH2}) and (\ref{eq:HX2}) by the coefficients
%%%%%%%%%%%%%%%%%%%%%%%%%%%%%%%%%%%%%%%%%%%%%%%%%%%%%%%%%%%%%%%%%%
\begin{equation}
\kappa_{2,4}^{(b)}(\lambda)\equiv
\kappa_{2,4}^{(b)}(\lambda;T,N_{a},N_{i})=K_{2,4}^{(b)}(\lambda,T)N_{a}N_{i},
\label{eq:kapa24}
\end{equation}
%%%%%%%%%%%%%%%%%%%%%%%%%%%%%%%%%%%%%%%%%%%%%%%%%%%%%%%%%%%%%%%%%%
where $T$, $N_{a}$, $N_{i}$ and $N_{mi}$ denote the local plasma temperature and the
densities of the atoms ($He$ or $H$), ions ($H^{+}$ or $X^{+}$) and molecular
ions ($HeH^{+}$ or $HeX^{+}$). Assuming the existence
of LTE, we will take the photo-dissociation coefficient
$\kappa_{1,3}^{(a)}(\lambda)$ in an equivalent form (suitable for
further considerations), given by the relations
%%%%%%%%%%%%%%%%%%%%%%%%%%%%%%%%%%%%%%%%%%%%%%%%%%%%%%%%%%%%%%%%%%
\begin{equation}
\kappa_{1,3}^{(a)}(\lambda)\equiv
\kappa_{1,3}^{(a)}(\lambda;T,N_{mi})=K_{1,3}^{(a)}(\lambda,T)N_{a}N_{i},
\label{eq:kapa13}
\end{equation}
%%%%%%%%%%%%%%%%%%%%%%%%%%%%%%%%%%%%%%%%%%%%%%%%%%%%%%%%%%%%%%%%%%
%%%%%%%%%%%%%%%%%%%%%%%%%%%%%%%%%%%%%%%%%%%%%%%%%%%%%%%%%%%%%%%%%%
\begin{equation}
K_{1,3}^{(a)}(\lambda,T)=\sigma_{1,3}^{phd}(\lambda,T) \cdot \chi_{1,3}^{-1}(T), \quad
\chi_{1,3}(T)=(N_{a}N_{i})/N_{mi},
\label{eq:chi}
\end{equation}
%%%%%%%%%%%%%%%%%%%%%%%%%%%%%%%%%%%%%%%%%%%%%%%%%%%%%%%%%%%%%%%%%%
where the photo-dissociation cross-section and the quantity
$\chi_{1,3}(T)$ are determined in a way similar to that in \cite{ign09}.

The efficiency of the processes (\ref{eq:HeH1}) and (\ref{eq:HeH2}), and (\ref{eq:HX1}) and
(\ref{eq:HX2}) together is characterized by the total spectral absorbtion coefficients
%%%%%%%%%%%%%%%%%%%%%%%%%%%%%%%%%%%%%%%%%%%%%%%%%%%%%%%%%%%%%%%%%%
\begin{equation}
\label{eq:kappatot}
\kappa_{12}(\lambda)=\kappa_{1}^{(a)}(\lambda)+\kappa_{2}^{(b)}(\lambda)=
K_{12}(\lambda,T)N_{a}N_{i},\quad
K_{12}(\lambda,T)=K_{1}^{(a)}(\lambda,T)+K_{2}^{(b)}(\lambda,T),
\end{equation}
%%%%%%%%%%%%%%%%%%%%%%%%%%%%%%%%%%%%%%%%%%%%%%%%%%%%%%%%%%%%%%%%%%
\begin{equation}
\label{eq:kappaab}
\kappa_{34}(\lambda)=\kappa_{3}^{(a)}(\lambda)+\kappa_{4}^{(b)}(\lambda)=
K_{34}(\lambda,T)N_{a}N_{i}, \quad
K_{34}(\lambda,T)=K_{3}^{(a)}(\lambda,T)+K_{4}^{(b)}(\lambda,T),
\end{equation}
%%%%%%%%%%%%%%%%%%%%%%%%%%%%%%%%%%%%%%%%%%%%%%%%%%%%%%%%%%%%%%%%%%
where $K_{1,3}^{(a)}(\lambda,T)$ is given by Eqs.~(\ref{eq:chi}).

\section{Results and Discussion}

The data needed for determination of $U_{1,2}(R)$ and $D_{12}^{2}(R)$ in the case $HeH^{+}$
are taken here from \cite{gre74a}.
For the molecular ions $HNa^{+}$ and $HLi^{+}$ the values of $U_{1,2}(R)$ and $D_{12}^{2}(R)$
are obtained here, using the method of calculation described in details in \citet{ign05}.

As the main aim of this work is to draw attention to the processes (\ref{eq:HeH1}) -
(\ref{eq:HX2}) as factors of influence on the opacity of the stellar atmospheres in UV
and VUV region, the spectral absorption coefficients (for Na, Ni =1) $K_{12}(\lambda,T)$ and
$K_{34}(\lambda,T)$, which characterize the total efficiency of the processes (\ref{eq:HeH1})
and (\ref{eq:HeH2}), and (\ref{eq:HX1}) and (\ref{eq:HX2}) respectively, will be determined.
The necessary calculations are performed here in accordance with (\ref{eq:kapa24}) -
(\ref{eq:kappaab})
in the part of the VUV region, namely $70 \textrm{nm} \le \lambda \le 170 \textrm{nm}$, for
several values of $T$
which are relevant for the photospheres of the Sun and of the considered DB white dwarfs.

The behavior of the total spectral absorption coefficients $K_{12}(\lambda,T)$ for
$70 \textrm{nm} \le \lambda \le 115 \textrm{nm}$ and $T=5500$K and $12000$K is shown in Figs.~1
and 2.
The first temperature is characteristical of a large part of the solar photosphere, and the
second - of a part of the atmosphere of DB white dwarfs with $T_{eff}=12000$K.

The behavior of the total spectral absorption coefficients $K_{34}(\lambda,T)$, in the
case of the processes (\ref{eq:HX1}) and (\ref{eq:HX2}) with $X=Na$, for
$130 \textrm{nm} \le \lambda \le 170 \textrm{nm}$ and $T=4000$K, is shown in Fig.~3. Let us note
that
the chosen temperature corresponds to a sunspot ( see \citet{mal86}).

Finally, taking into account the existence of the so-called Lithium stars, intensively discussed
in literature \citep[see][]{nor98,sha03}, in Fig.~4 the behavior is shown of the total spectral
absorption coefficients $K_{34}(\lambda,T)$, in the case of the processes (\ref{eq:HX1}) and
(\ref{eq:HX2}) with $X=Li$, for the same $\lambda$ and $T$ as in the case $X=Na$.

The presented figures illustrate the fact that the considered non-symmetric processes
(\ref{eq:HeH1}) - (\ref{eq:HX2}) generate rather wide molecular absorption bands in the VUV
region.
Moreover, the comparison of $K_{12}(\lambda,T)$ and $K_{34}(\lambda,T)$
with the corresponding characteristics of other relevant absorption processes (symmetric
ion-atom absorption processes and etc.), as well as the corresponding particle densities,
suggest a conclusion that in the regions of $\lambda$ which correspond to the most parts of the
mentioned bands the efficiency of the processes (\ref{eq:HeH1}) - (\ref{eq:HX2}) is at least
not negligible, and sometimes it is just these processes that determine the opacity of the
considered stelar atmospheres.

%%%%%%%%%%%%%%%%%%%%%%%%%%%%%%%%%%
\textbf{Acknowledgments} This work was supported by the Ministry of
Education, Science and Technological Development of the Republic of Serbia as a part of the
projects 176002, III4402.

%\bibliographystyle{apa}
%\bibliography{b}

\newcommand{\noopsort}[1]{} \newcommand{\printfirst}[2]{#1}
  \newcommand{\singleletter}[1]{#1} \newcommand{\switchargs}[2]{#2#1}

\begin{figure}[ht]
\begin{minipage}[b]{0.5\linewidth}
\centering
\includegraphics[scale=0.25]{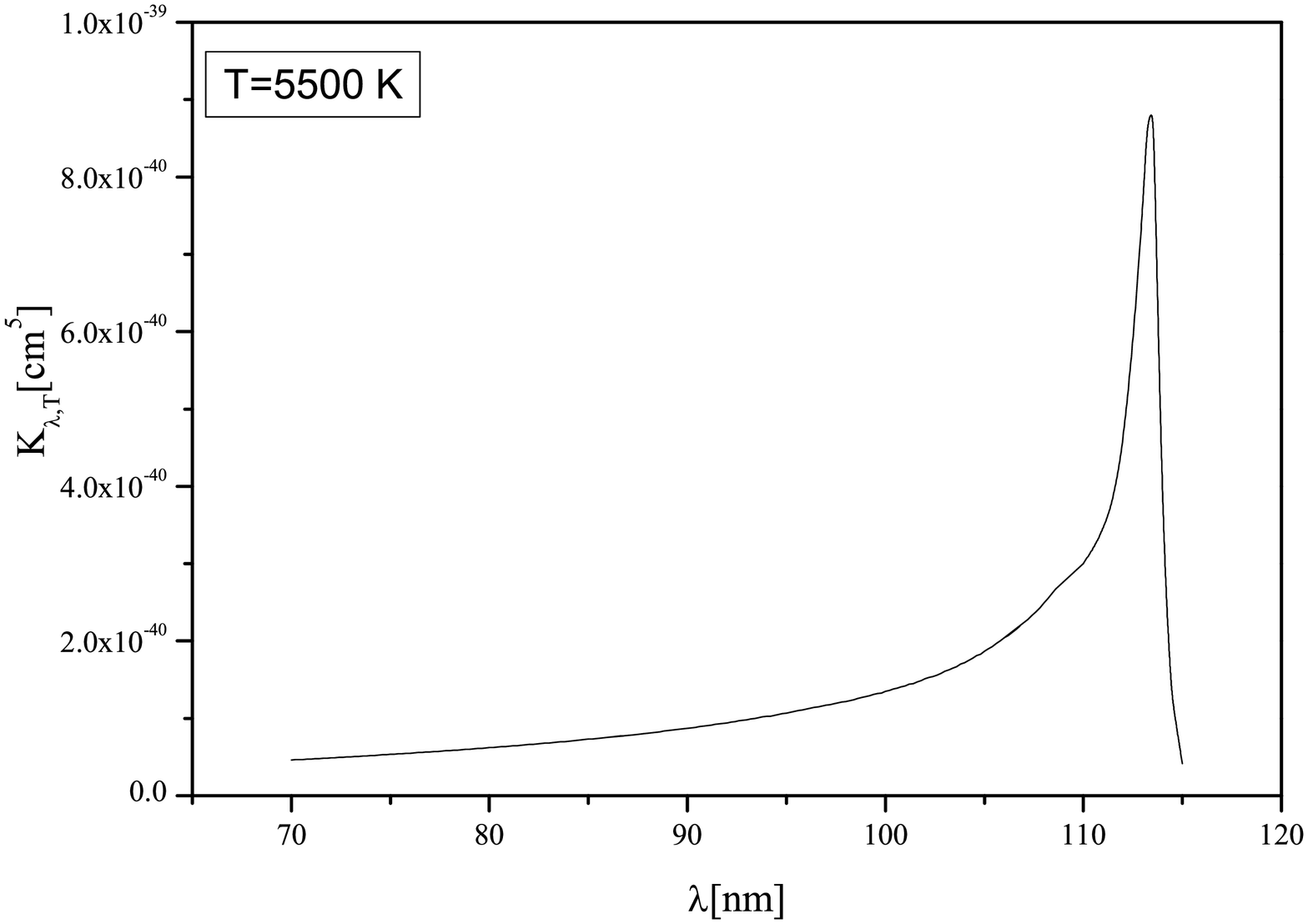}
\caption{Rate coefficients
$K_{ia}(\lambda,T)$ (spectral absorption coefficients for Na, Ni =1) at T=5500K, for $HeH^{+}$.}
\label{fig:figure7}
\end{minipage}
\hspace{0.5cm}
\begin{minipage}[b]{0.5\linewidth}
\centering
\includegraphics[scale=0.25]{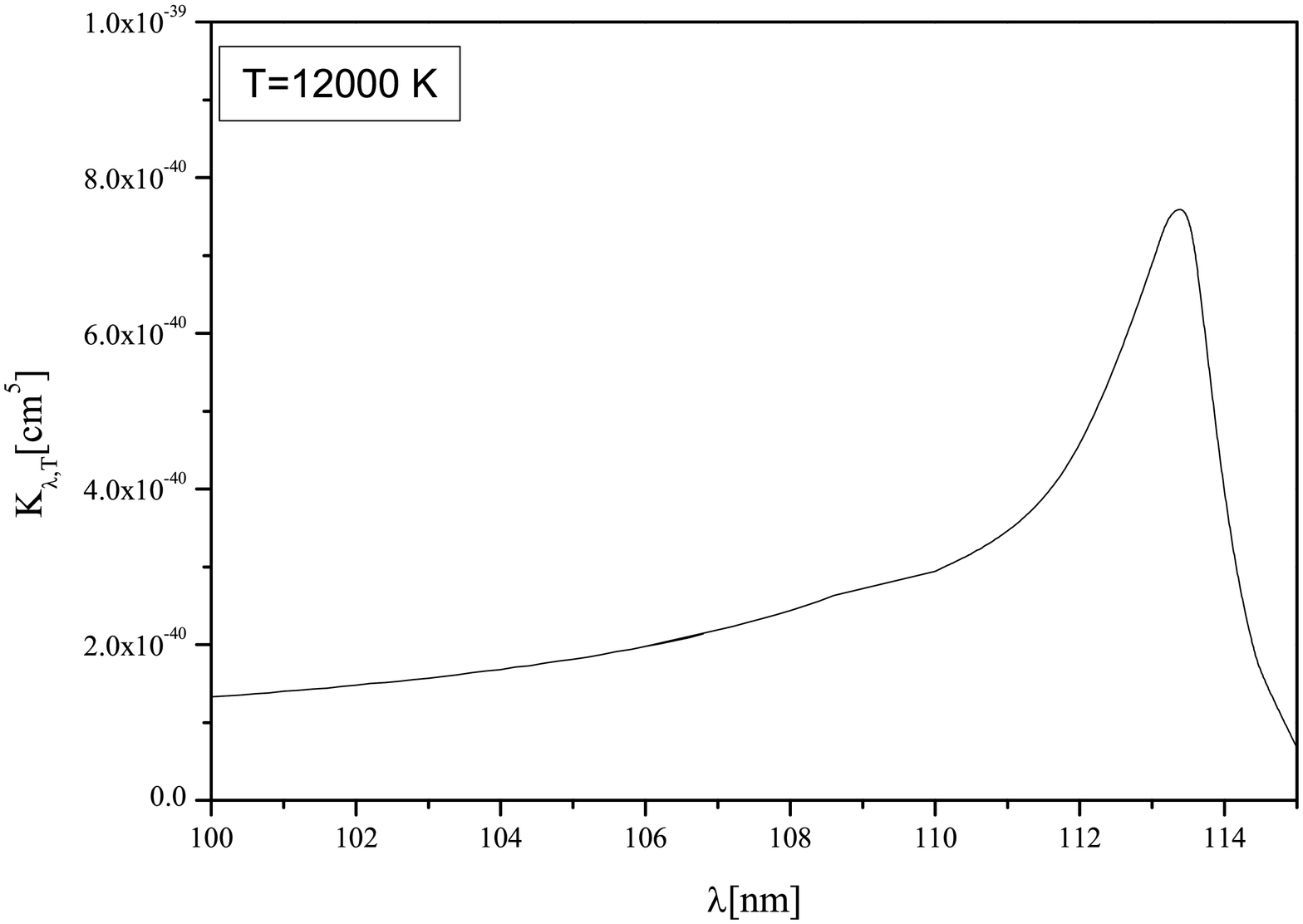}
\caption{Rate coefficients
$K_{ia}(\lambda,T)$ at T=12000K, for $HeH^{+}$.}
\label{fig:figure8}
\end{minipage}
\end{figure}

\begin{figure}[ht]
\begin{minipage}[b]{0.5\linewidth}
\centering
\includegraphics[scale=0.25]{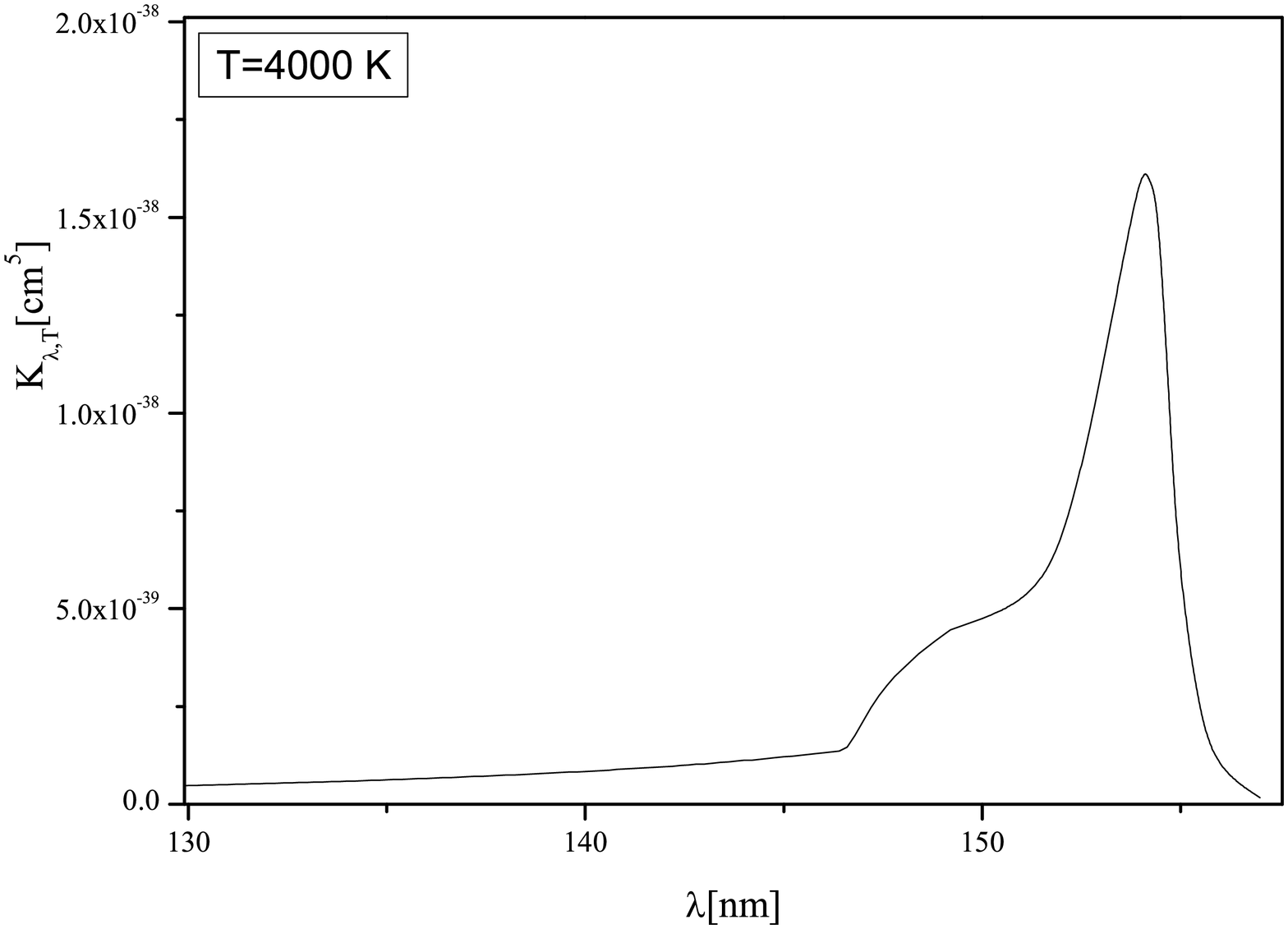}
\caption{Rate coefficients
$K_{ia}(\lambda,T)$ at T=4000K, for $HNa^{+}$.}
\label{fig:figure9}
\end{minipage}
\hspace{0.5cm}
\begin{minipage}[b]{0.5\linewidth}
\centering
\includegraphics[scale=0.25]{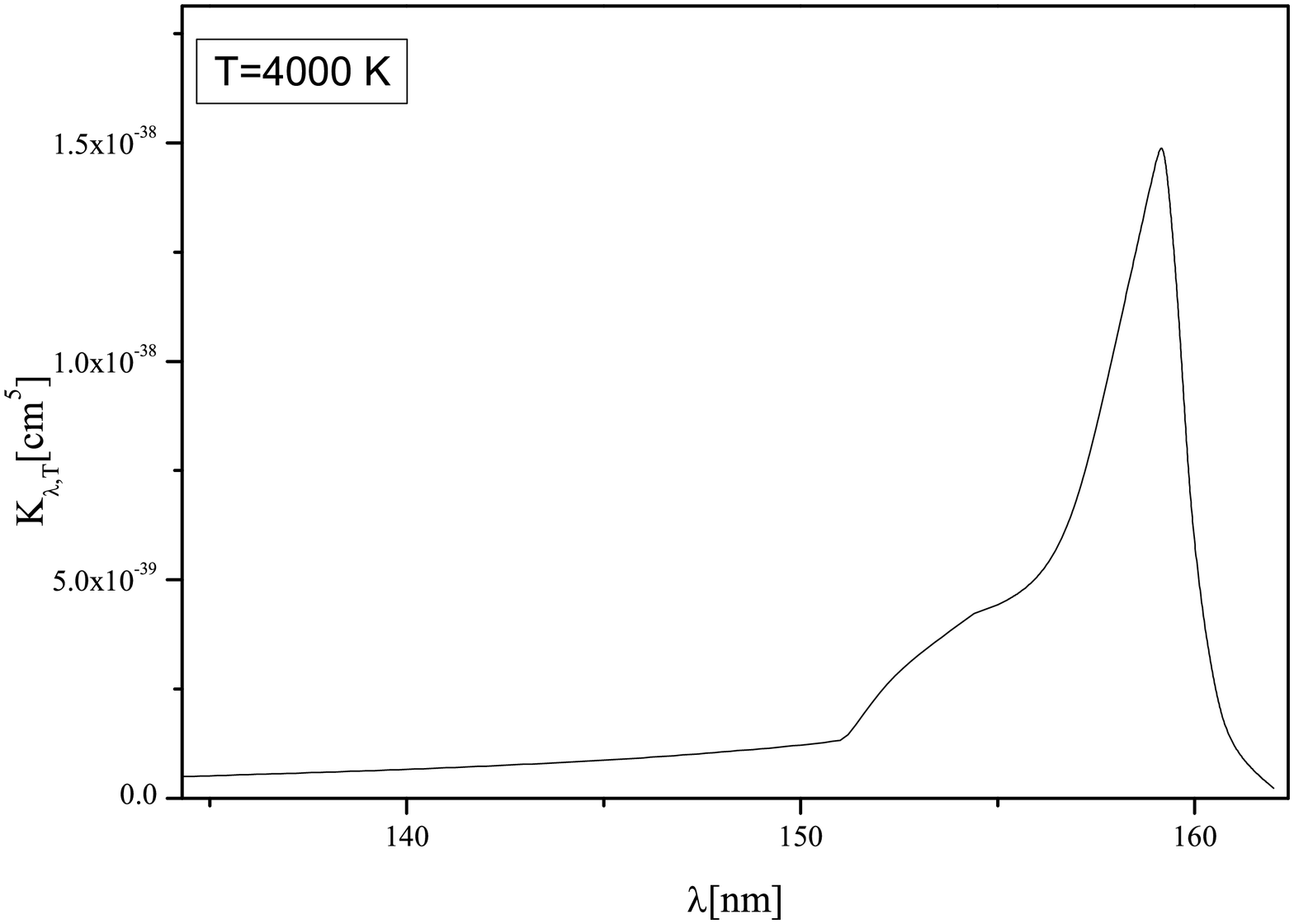}
\caption{Rate coefficients
$K_{ia}(\lambda,T)$ at T=4000K, for $HLi^{+}$.}
\label{fig:figure10}
\end{minipage}
\end{figure}

\end{document}